\documentclass[aps,pra,twocolumn,floatfix]{revtex4-1}
\usepackage{graphicx}
\usepackage{amsmath}
\usepackage{braket}
\usepackage{tikz}
\begin{document}

\title{Stable production of a strongly-interacting Bose-Einstein condensate via mode-matching}
\begin{abstract}
We describe a diabatic protocol to prepare a strongly-interacting Bose-Einstein condensate in a regime where neither an adiabatic ramp nor a direct diabatic quench are desirable. This protocol is expected to achieve a nearly unit population transfer to the strongly-interacting ground state for realistic experimental parameters for $^{85}$Rb. The protocol matches the initial and final density profiles by modifying the trap along with the scattering length during the quench. The protocol should reveal several properties of the strongly-interacting Bose gas, and enable further investigation of beyond mean-field corrections to the Gross-Pitaevskii equation. 
\end{abstract}

\author{E. J. Halperin}
\email{eli.halperin@colorado.edu}
\affiliation{JILA, NIST, and Department of Physics, University of Colorado, Boulder, Colorado 80309-0440, USA}
\author{M. W. C. Sze}
\affiliation{JILA, NIST, and Department of Physics, University of Colorado, Boulder, Colorado 80309-0440, USA}
\author{J. P. Corson}
\affiliation{JILA, NIST, and Department of Physics, University of Colorado, Boulder, Colorado 80309-0440, USA}
\author{J. L. Bohn}
\affiliation{JILA, NIST, and Department of Physics, University of Colorado, Boulder, Colorado 80309-0440, USA}
\date{\today}
\maketitle

\section{Introduction}

An outstanding goal in the physics of dilute, atomic Bose-Einstein condensates (BECs) is to track their properties from the weakly-interacting regime to the strongly-interacting regime. For a dilute gas, whose mean interatomic spacing is far greater than the range of the two-body interaction, the strength of interaction is conveniently parametrized by the two-body scattering length $a$, which can be tuned by means of magnetic field Fano-Feshbach resonances. If the scattering length and the number density $n$ satisfy $na^3 \ll 1$, then $na^3$ is a perturbative parameter, and the usual methods of quantum field theory describe the physics quite accurately. However, $a$ can also be set to large, even infinite values, which calls for a re-appraisal of how these Bose gases behave. 

A presumptive experimental procedure for producing a ground state BEC at large scattering length would start with a stable ground state BEC at some small initial scattering length $a_i$. The scattering length would then be ramped adiabatically to the desired final value $a_f$, slowly enough that the gas remains in the ground state of the trap   throughout the ramp. This effort is limited, however, by the depletion and heating of the condensate caused by three-body recombination, which scales as the fourth power of the scattering length. These circumstances constrain the speed of the adiabatic ramp, for it must be fast enough to not lose too many atoms but slow enough to track the ground state along the way. 

For sufficiently small scattering lengths, an adiabatic ramp of $a$ is perfectly reasonable, as in the experiment of Ref.~\cite{navon2011dynamics}. This experiment measured the beyond mean-field correction predicted by Lee, Huang, and Yang (LHY)~\cite{LHY1, LHY2, dalfovo1999theory} to the equation of state up to moderate scattering length of $2000 a_0$, where the perturbative parameter $n a^3 \approx 0.049$. However, the larger the final scattering length, the more impractical an adiabatic ramp becomes. One partial solution to this difficulty is to study the large-$a$ thermal Bose gas, where the destructive effects of three-body recombination are minimized~\cite{fletcher2013stability,rem2013lifetime}. Another is to make the transition $a_i \rightarrow a_f$ rapidly and to study the dynamics before the gas is destroyed. This is in fact the technique used to explore the unitary BEC, where the scattering length is infinite~\cite{smith2012condensation, fletcher2013stability, makotyn2014universal}. 

This kind of diabatic quench of $a$ is not without its problems. Chief among them is that the initial and final ground states are not well matched, as illustrated in Fig. 1(a). This sketch includes a representation of a harmonic trap with frequency $\omega_f$ (blue curves), in which the strongly-interacting BEC is prepared. The density profile of the BEC is also sketched (black lines) for the initial scattering length $a_i$ in the lower panel, and the final $a_f$ in the upper panel. The density profile of $a_f$ is somewhat wider if $a_f>a_i$. It is evident that, under a diabatic quench that projects the $a_i$ wave function onto the $a_f$ wave function, the condensate's immediate response is to expand rapidly. It is not in its stationary ground state, and is moreover at a higher density than the ground state, exacerbating the three-body loss.

 \begin{figure}
  \centering
  \begin{tikzpicture}
    \node at (-4,-6) {\large (a)};
    \draw[blue,thick,opacity=0.7] (-2,-4) parabola bend (0,-5) (2,-4) node[below right] {};
    \draw[blue,thick,opacity=0.7] (-1,-1) parabola bend (0,-3) (1,-1) node[below right] {};
    \draw[dashed,thick,opacity=0.3] (-0.5,-5) -- (-0.5,-2);
    \draw[dashed,thick,opacity=0.3] (0.5,-5) -- (0.5,-2);

    \draw[thick] (-0.5,-5) parabola bend (0,-4.5) (0.5,-5) node[below right] {};
    \draw[thick] (-0.5,-2.6) parabola bend (0,-2.1) (0.5,-2.6) node[below right] {};

    \draw[thick,,->,>=stealth] (1.2,-4) -- (1.2,-1.7) node[midway,right] {a};

    \draw[thick,->,>=stealth] (1.9,-4) -- (1.9,-1.7) node[midway, right] {$\omega$};

    \node at (0,-6) {\large (b)};
    \draw[blue,thick,opacity=0.7] (-5,-1) parabola bend (-4,-3) (-3,-1) node[below right] {};
    \draw[blue,thick,opacity=0.7] (-5,-3) parabola bend (-4,-5) (-3,-3) node[below right] {};
    
    \draw[thick] (-4.2,-5) parabola bend (-4,-4) (-3.8,-5) node[below right] {};
    \draw[thick] (-4.5,-2.6) parabola bend (-4,-2.1) (-3.5,-2.6) node[below right] {};
    \draw[dashed,thick,opacity=0.3] (-4.2,-5) -- (-4.2,-2);
    \draw[dashed,thick,opacity=0.3] (-3.8,-5) -- (-3.8,-2);

    \draw[thick,,->,>=stealth] (-5.3,-4) -- (-5.3,-1.7) node[midway,left] {a};
    
  \end{tikzpicture}
  \caption{A schematic drawing of (a) the single quench of scattering length in a given trap, versus (b) the simultaneous double quench that projects both scattering length and trap frequency to their final values.}
  \label{schematic}
\end{figure}
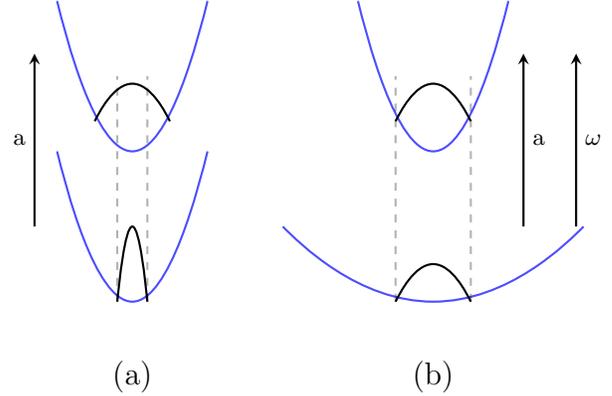

As an alternative, we here propose a mode-matching protocol to offset the effect of jumping to the strongly-interacting regime by also modifying the trap shape during the quench. This protocol is sketched in Fig. 1(b). We aim to produce a BEC with scattering length $a_f$ in a trap with frequency $\omega_f$, i.e., the upper panel of Fig.~1(b) presents the same goal as the upper panel of Fig.~1(a). In the stable initial state with low scattering length $a_i$, we instead begin in a larger trap where the initial wave function is the same width as the final target BEC (lower panel of Fig.~1(b)). We then simultaneously quench both the trap to its final frequency and the scattering length to its final value. By means of this double quench, one can prepare a strongly-interacting gas whose density profile closely resembles that of the BEC in its ground state. 
 
In a harmonic trap the initial frequency $\omega_i$ required to achieve this protocol is easily estimated in the Thomas-Fermi (TF) approximation. In this case the radius of the BEC in a spherically symmetric trap is given by~\cite{dalfovo1999theory}
\begin{align}
 R_{TF} = a_{ho} \left( \frac{15Na}{a_{ho}} \right)^{1/5},
\end{align}
where $a_{ho} = \sqrt{ \hbar /m \omega}$ is the oscillator length, $N$ is the total number of atoms, and $a$ is the scattering length. Since both traps are harmonic, two states with equal TF radii will have exactly equal density profiles in this approximation, even when the trap frequencies are different. For a given final trap frequency $\omega_f$, final scattering length $a_f$, and initial scattering length $a_i$, the optimal initial trap frequency $ \omega_{TF}$ is therefore
\begin{align}
 \omega_{TF} = \sqrt{\frac{a_i}{a_f}} \omega_f .
\label{TFomega}
\end{align}

In this approximation, the ground state wave function for the initial trap, $\phi_i$,  has perfect overlap with the ground state wave function for the strongly-interacting gas, in the final trap $\phi_f$, and would therefore produce the ground state of the strongly-interacting BEC exactly. This is in contrast to the direct quench of scattering length where the trap frequency is the same before and after the quench. Upon the direct quench the overlap would instead be

\begin{align}
 \left| \braket{\phi_i | \phi_f} \right|^2 \approx \frac{225 \pi^2}{1024} \left(\frac{a_i}{a_f} \right)^{3/5}.
 \label{approxdirectquench}
\end{align}
This is valid when $a_f \gg a_i$, since we approximated the final density as constant where the initial density is nonzero. This gives about $21\%$ overlap for the case considered in detail in the next section.

 Even if the densities before and after the quench are perfectly matched, there may still be correlations created during the quench, leading to quasiparticle excitations in the final state~\cite{stefan2013}. In order to reach the ground state of the strongly interacting gas exactly, the quench, which is diabatic on the time scale of trap dynamics, must be adiabatic on the time scale of quasiparticle excitations and molecular dynamics. The characteristic time scale on which quasiparticle excitations develop is $\tau_q = \hbar/(g n)$. Thus, for a spherically symmetric trap in the TF approximation
 \begin{align}
     \tau_q \approx \frac{1}{15^{2/5} \pi} \frac{a_{ho}}{N a} \tau_\omega,
 \end{align}
where $\tau_\omega = 2\pi/\omega$ is the time scale of the trap. For parameters given in the next section, $\tau_q = 0.003 \tau_\omega$ in the final trap. Likewise, molecular dynamics~\cite{corson2016ballistic,corson2015,Hung1213} occur on a time scale $\tau_m = m a^2/\hbar$, where for our parameters $\tau_m = 0.001 \tau_\omega$. Since these time scales are both much faster than the trap dynamics, we approximate the quench as adiabatic for any two-body and molecular dynamics, but diabatic for the density dynamics. Consequently, we ignore two-body dynamics in what follows. 

The true BEC is described by physics not included in the Thomas-Fermi approximation, of course. In this paper, we evaluate the protocol for more realistic BECs that are described by physics including a correction beyond the mean-field approximation, and that incorporate realistic losses due to three-body recombination. In general, in projecting from a harmonic trap to another harmonic trap, the overlap of initial and final wave functions will not be exactly perfect, and the resulting BEC will contain some residual radial oscillations. Therefore, an operational goal of the mode-matching procedure is to produce a gas with the minimum residual breathing excitation -- a goal that is generally achievable for an arbitrary final state. In cases where the final state is expected to have a known profile, we can improve the method and suggest a modified initial potential, not necessarily harmonic, that guarantees unit overlap.

\section{State Preparation by the Mode-Matching Protocol}

\begin{table}
  \begin{tabular}{llcr}
  \toprule
  Number of atoms                 & $N$   & $=$ & $10^5$ \\
  Initial scattering length       & $a_i$ & $=$ & $100a_0$ \\
  Final scattering length         & $a_f$ & $=$ & $5000a_0$ \\
  Final trap frequency            & $\omega_f$ & $=$ & $2 \pi \times 10$ Hz\\
  Loss coefficient                     & $K_3$      & $=$ & $2 \times 10^{-23}$ cm$^6$/s\\
  Perturbative parameter        & $na^3$ & $=$ & $0.025$ \\
  \botrule
  \end{tabular}
  \caption{System parameters for the double quench modeled in Figures 2-4.}
  \label{table:params}
\end{table}

We model a BEC containing $N$ atoms in a spherically symmetric harmonic oscillator trap with frequency $\omega$ and scattering length $a$. Throughout, we model the dynamics of the condensate using a modified Gross-Pitaevkii equation~\cite{LHY1,LHY2,dalfovo1999theory}
\begin{align}
i\hbar\frac{\partial\Psi}{\partial t} = \left[-\frac{\hbar^2 \nabla^2}{2m}+ V + g\vert\Psi\vert^2 + g'\vert\Psi\vert^3 - i \hbar \frac{K_3}{2} \vert \Psi \vert^4 \right]\Psi ,
\label{GPE}
\end{align}
where $\Psi$ is the many-body order parameter of the Bose gas, realted to the single particle wave function $\phi$ by $\Psi = \sqrt{N} \phi$. Here $V(r) = m \omega^2 r^2/2$ is the external potential, and $g= 4\pi \hbar^2 a/m$ is the coupling constant, with $a$ the two-body scattering length. The LHY term is given by the coefficient $g^\prime=32/(3\sqrt{\pi})a^{3/2}g$, and losses are incorporated via the three-body loss coefficient $K_3$ which is a function of the scattering length and the atomic species and is defined by the loss it incurs,
\begin{equation}
 \frac{d}{dt} n = -K_3 n^3 .
\end{equation}
$K_3$ has two components, for a deep quench $K_d$ and a shallow quench $K_s$, depending on the molecular state produced in the recombination. The total loss rate is $K_3= K_{d} + K_{s}$, where \cite{braaten2006universality}
\begin{align}
 K_{d} &= 33.4\left( 1 - e^{-4\eta} \right) \frac{\hbar a^4}{m}, \\
 K_{s} &= 134.2 e^{-2\eta} \left( \sin^2 \left( s_0 \ln\left(a\kappa \right) +4.3\right) + \sinh^2\eta\right) \frac{\hbar a^4}{m}, \nonumber
 \label{eq:loss}
\end{align}
 with $s_0 = 1.006$. For  $^{85}$Rb, $\eta=0.057(2)$ and $\kappa=39(1)$~$\mu m^{-1}$~\cite{wild2012measurements}. Thus, $K_3 = 2 \times 10^{-23}$ cm$^6$/s for scattering length $a_f = 5000a_0$ used in the calculations below. Our perturbative parameter $na^3 = 0.025$, although the LHY correction to the chemical potential is $\sqrt{\pi n a^3} = 0.280$. We are in a regime where the LHY correction is appreciable but the next order term can be mostly neglected. 
 
\subsection{Harmonic Trap}
\label{harmonicTrap}

\begin{figure}[tb]
  \centering
    \includegraphics[width = 0.48\textwidth]{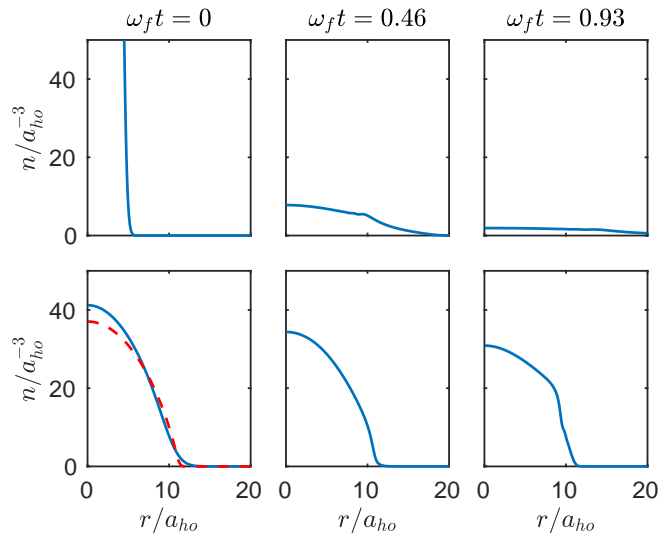}
    \caption{Snapshots of time evolution of the direct quench (top) versus the double quench (bottom). In both cases the final BEC is specified by the parameters in Table I. In the upper panel the initial trap frequency is $\omega_i = \omega_f = 2 \pi \times 10$ Hz, while in the lower panel it is $\omega_i = 2 \pi \times 1.1$ Hz.}
    \label{timeEvo}
\end{figure}

We first contrast the mode-matching protocol to the direct scattering length quench in Fig.~\ref{timeEvo}. The first row depicts the direct quench of the scattering length from $a_i=100a_0$ to $a_f=5000a_0$, in a fixed trap of frequency $\omega_f=2 \pi \times 10$ Hz. This trap was chosen to emulate the trap used in the unitary gas experiments at JILA~\cite{makotyn2014universal}. The density profile is shown at three time intervals after the quench. Upon suffering the direct quench, the gas rapidly expands and loses atom number due to its high initial density. Moreover, the density profile is no longer smooth.

The second row in Fig.~\ref{timeEvo} shows the density profile at the same times, for a protocol that also includes a quench from $\omega_i = 2 \pi \times 1.1 $~Hz to $\omega_f = 2 \pi \times 10$ Hz along with the scattering length quench. This initial frequency is chosen to minimize the breathing excitations after the quench, in a way to be specified below. This double quench preserves the general shape of the density profile. Still, the gas is naturally densest near the center of the trap where $r=0$. Thus, there is more loss near the center, which distorts the density profile over time. The gas shrinks when the central density decreases, approximately tracking whatever shape the ground state density profile with fewer atoms would take. This process, however, is not adiabatic and causes a slight wobble, as seen in the bottom right panel.

 The mode-matching double quench also has the advantage of starting at a lower density than the direct quench, and thus preserving more atoms in the BEC. At $\omega_ft = 0.46$, the gas retains only around $57\%$ of the original atoms in the direct quench, whereas the gentler double quench retains around $95\%$ of the original atoms. 

It is worth noting that the state produced in the double quench is not the same as the true final ground state (red dashed curve). Although this slight mismatch is unavoidable for a harmonic initial trap, the initial frequency is chosen to minimize oscillatory dynamics, not maximize overlap with the ground state. The optimal initial frequency is $\omega_{opt} = 0.78 \omega_{TF} = 2 \pi \times 1.1$ Hz, instead of $0.85 \omega_{TF} = 2 \pi \times 1.2$~Hz, the frequency which maximizes overlap onto the ground state. The overlaps in the two cases are $\left| \braket{\phi_i | \phi_f} \right|^2 = 0.975$  and $0.980$, respectively. 

\begin{figure}[tb]
  \centering
    \includegraphics[width = 0.48\textwidth]{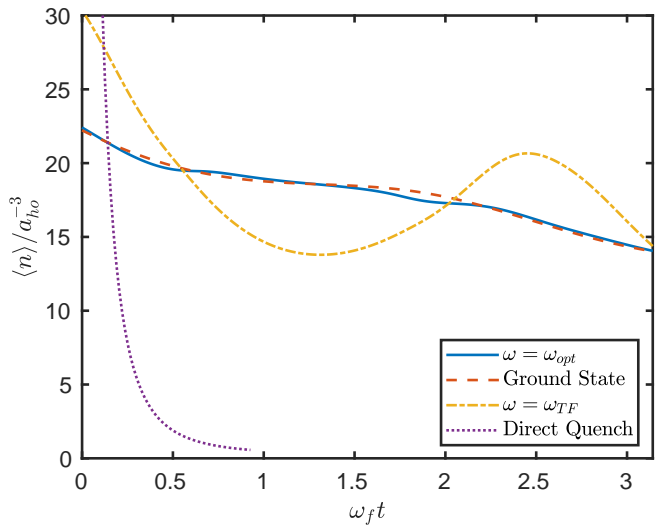}
    \caption{Mean density over one half the trap period. The direct quench is only shown for a fraction of the trap period, due to the model's difficulty accurately describing the haphazard dynamics of the direct quench at long times.}
    \label{densityDynamics}
\end{figure}
The dynamical response of the BEC created in the double quench is a sensitive function of the initial trap frequency $\omega_i$. This is illustrated in Fig.~\ref{densityDynamics}, which shows the time evolution of the mean density $\braket{n(t)}$, defined as
\begin{align}
    \braket{n(t)} = \frac{4\pi}{N} \int_0^\infty r^2 n^2(t) dr .
\end{align}
This figure shows $\braket{n(t)}$ for four different initial conditions before the quench. The dotted purple curve corresponds to the direct quench with $\omega_i=\omega_f$, and shows the expected, rapid decrease of the density. Next, the dashed-dot yellow line gives the response if we had selected the initial frequency $\omega_i = \omega_{TF}$ according to the TF criterion in Eq.~(\ref{TFomega}). This value of $\omega_i$ leads to a significantly different density profile from the gas prepared with $\omega_i = \omega_{opt}$. Here the overlap squared is $\left| \braket{\phi_i | \phi_f} \right|^2 = 0.961$. The gas is denser in the center than the true ground state $\phi_f$, and so it expands and then contracts, showing a breathing mode around some overall decaying density. These oscillations are amplified by the nonlinearity of Eq.~\eqref{GPE} but damped by the loss. Generally, oscillations lead to an overall decrease in atom number over one oscillatory period as compared to pure exponential decay, as the loss is proportional to $n^3$~\cite{braaten2006universality}. Thus, the initial state that minimizes oscillatory dynamics also maximizes the number of atoms still remaining after one oscillatory period.

The solid blue line in Fig.~\ref{densityDynamics} gives the mean density for the optimal double quench, chosen to minimize the oscillations in $\langle n(t) \rangle$ around an overall decay. The optimal frequency gives very similar density dynamics to the BEC that starts in the true ground state of the final trap (red dashed line), which is unsurprising due to their high overlap. While the double quench does not produce the exact ground state of the strongly-interacting gas, it nevertheless produces something very similar. 

The behavior of the mean density over time suggests an operational procedure for optimizing the initial frequency. Dynamics that do not distort the shape of the density can be approximated by exponential decay of $\braket{n(t)}$. Dynamics beyond this exponential decay are due to the mode mismatch or loss, both of which contribute to radial breathing modes. We can effectively use the amplitude of the oscillations around the exponential profile as a measure of our mode-matching. We first find the best fit to the mean density for a decaying exponential,
\begin{align}
    f(t) = f_0 \exp (-\gamma t ).
\end{align}
We define the mode mismatch $\Delta$, measuring the magnitude of the oscillatory motion, as the mean squared error of the actual density profile from the best fit curve, given by 
\begin{align}
    \Delta = \frac{1}{\tau}\int_0^\tau (f(t) - n(t))^2 dt ,
\end{align}
where $\tau$ is the time for which we observe the dynamics, in this case one oscillatory period. This $\Delta$ is a measure of how well the mode produced in the double quench matches the sought ground state. Unlike overlap, this is a model independent and experimentally observable quantity.

\begin{figure}[tb]
  \centering
    \includegraphics[width = 0.48\textwidth]{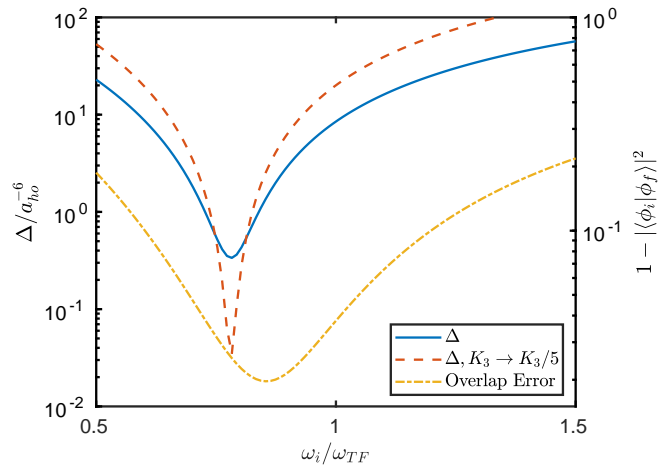}
    \caption{Mode mismatch $\Delta$ as functions of initial trap frequency. The blue curve shows the nominal result for the model specified in Table I; the minimum of this curve is used to determine the optimum initial frequency for the double quench. The red dashed curve is the same, but for an artificially reduced three-body loss rate. For comparison, the dotted line shows the direct overlap of the initial and final wave functions.}
    \label{spectroscopy}
\end{figure}

Figure \ref{spectroscopy} plots $\Delta$ over a range of initial frequencies for two different values of the loss parameter (red and blue curves). The minimum of $\Delta$ was used to determine $\omega_{opt} = 2 \pi \times 1.1 $~Hz, the initial trap frequency for Fig.~\ref{timeEvo} and Fig.~\ref{densityDynamics}. As a reference, we also show the overlap error (yellow dotted curve), defined as $1 - \left| \braket{\phi_i | \phi_f} \right|^2$. The location of the minimum is shifted for $\Delta$ as compared to the minimum of overlap error. This leads to a slightly larger initial gas, offsetting the wobble caused by increased loss near the center of the trap. Since overlap, unlike $\Delta$, is not a dynamical quantity, it is independent of loss and thus only plotted once. For comparison, the red dashed curve shows the same $\Delta$ for a hypothetical case where the loss coefficient $K_3$ is five times smaller. Reducing the amount of loss leads to a sharpening of the $\Delta$ peak.
\subsection{Exact Preparation of the Ground State}
\label{trapshaping}
The mode-matching scheme is quite general. A double quench like that described above should produce a final-state BEC close to the desired ground state, even when that ground state is not known ahead of time. However, in a moderately-interacting regime where the final ground state is expected to be known, a modification of the scheme should allow one to reproduce it exactly. 

 Doing so  could extend the experimentally feasible regime beyond the $2000a_0$ limit in Ref.~\cite{navon2011dynamics}, where the adiabatic ramp is limited by three-body loss. Just beyond this regime, loss is still slow enough that any dynamics would be initially dominated by the mismatch onto the final ground state. Scanning the initial trap frequency and minimizing $\Delta$ improves the overlap over the direct quench, but still only gives an overlap squared of $0.975$ for our system parameters. Here we show how to eliminate the remaining discrepancy in projecting onto the known strongly-interacting state.

Given a target wave function that is the ground state $\phi_t$ of the final trap,  we seek a weakly-interacting state with perfect overlap onto $\phi_t$. This increases the complexity of the initial trap, which will no longer be harmonic, but will require the tools of trap shaping~\cite{zupancic2016ultra}. We design a potential $V$ such that the Gross-Pitaevskii equation with small scattering length has $\phi_t$ as its ground state. In this way $\phi_t$ could be adiabatically prepared in $V$ prior to some quench into large scattering length. At low scattering length, beyond LHY corrections and loss are negligible, so the Gross-Pitaevskii equation for the initial single-particle ground state wave function $\phi_i$ reads
\begin{align}
\left(-\frac{\hbar^2}{2m}\nabla^2+ V - \mu + g N \vert\phi_i\vert^2 \right)\phi_i = 0,
\label{newGPE}
\end{align}
where the chemical potential $\mu$ and the coupling $g = 4\pi \hbar^2 a_i/(2m)$ are both for the initial, low scattering length state. We set the potential $V$ based on the target state $\phi_t$ to 
\begin{align}
V = \frac{\hbar^2}{2m} \frac{\nabla^2 \phi_t}{\phi_t} - g N \vert\phi_t\vert^2 + \mu,
\label{Vsolve}
\end{align}
where $g$ and $\mu$ are still for the initial state. Inserting $V$ into Eq.~\eqref{newGPE} and rearranging terms, we have 
\begin{align}
\left[\frac{\hbar^2}{2m}\left(-\nabla^2 +  \frac{\nabla^2 \phi_t}{\phi_t}\right)  + g N \left(\vert\phi_i\vert^2  - \vert\phi_t\vert^2  \right) \right]\phi_i = 0,
\label{GPEdemo}
\end{align}

\begin{figure}[bt]
  \centering
    \includegraphics[width = 0.48\textwidth]{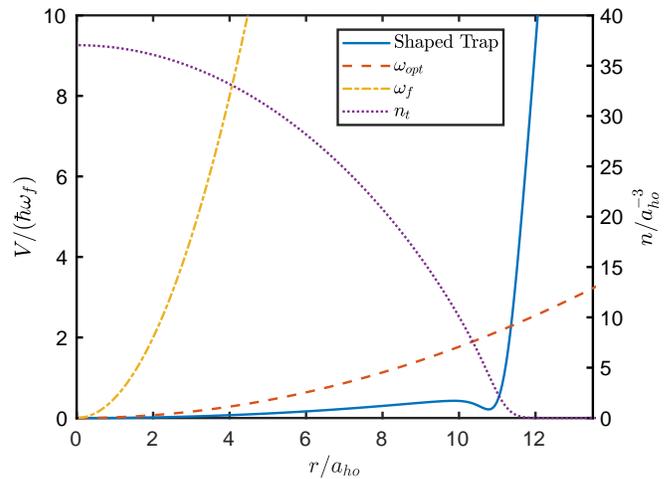}
    \caption{The ground state density and the shaped trap that gives this same density at low scattering length. Both the steep $\omega_f$ and the shaped trap give the same density, the former at $5000a_0$ and the latter at $100 a_0$. The chemical potential is significantly higher at $5000a_0$ than $100a_0$.}
    \label{ShapedV}
\end{figure}

Now $\phi_t$ solves Eq.~\eqref{GPEdemo}, since both terms are zero by setting $\phi_i = \phi_t$. This is an equation for the gas at low scattering length, although $\phi_t$ is the product of the high scattering length gas. Since the target state is the ground state of the final potential, it is real and nodeless~\cite{pítajevskii2003bose}. Thus, $\nabla^2 \phi_t/\phi_t$ is well behaved. The chemical potential sets an overall but irrelevant shift to $V$ in Eq.~\eqref{Vsolve}, so for convenience we set $V(r=0) = 0$.

For ground state $\phi_t$ of the harmonic trap at $a_f = 5000a_0$, we show the potential that produces this same state with $a_i = 100a_0$ in Fig.~\ref{ShapedV}. The shaped trap and optimal harmonic trap are both very wide on the scale of the final trap, as the scattering length is much smaller during the preparation than after the quench. However, the shaped trap that produces $\phi_t$ exactly is strikingly different from the optimal harmonic trap, especially given that we achieved overlap squared of $0.975$ (instead of unity) with the harmonic trap. The shaped trap rises much more quickly than the harmonic trap after $R_{TF} = 10.3$~$a_{ho}$, reducing the exponential tail of the density profile to better match the short tail of the strongly-interacting gas. Additionally, it stays below the harmonic trap, and is essentially flat until $r \approx R_{TF}$. For $r \gg R_{TF}$, both $\nabla^2 \phi_t \approx 0$ and $\phi_t \approx 0$, so their quotient is indeterminable. We therefore cut off the shaped trap at large $r$, well after the density has gone to zero. We numerically verify, using the full model including the LHY term, that this shaped potential gives the correct ground state wave function with overlap error less than $10^{-3}$.

\section{Ancillary Measurements of Gas Properties}
The mode-matching protocol, in addition to preparing an approximate or exact strongly-interacting ground state, offers a way to extract useful information about the gas and the equation of state that governs it. We demonstrate a method for observing the LHY correction and beyond LHY correction to the Gross-Pitaevskii equation, and how to approximately gauge the ground state density and energy.

\begin{figure}[bt]
  \centering
    \includegraphics[width = 0.48\textwidth]{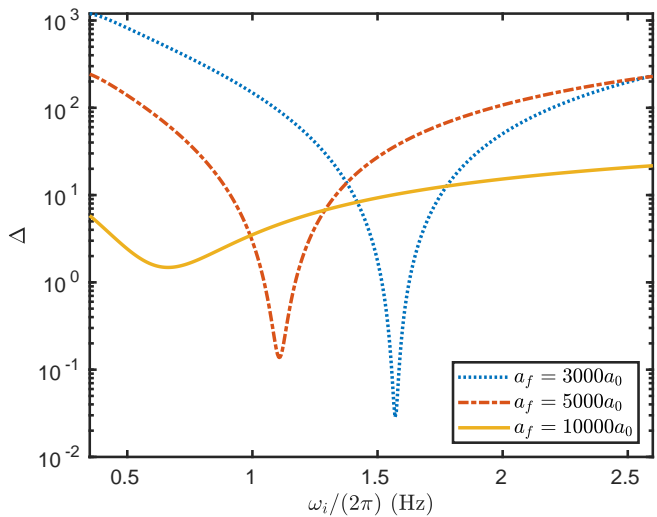}
    \caption{Mode mismatch $\Delta$ as a function of initial trap frequency $\omega_i$ for three final scattering lengths $a_f$. We move from the mean-field regime at $a_f = 3000a_0$ to the intermediate regime described throughout at $a_f = 5000a_0$, and finally to the beyond LHY regime at $a_f = 10000a_0$.}
    \label{3spectra}
\end{figure}

By considering the map of $\Delta$ versus $\omega_i$ as a kind of spectrum, one can extract the ``resonant'' frequency $\omega_{opt}$ at the minimal value of $\Delta$. Figure~\ref{3spectra} shows scans of $\omega_i$ for three different final scattering lengths $a_f$. We give these frequencies in units of Hz here instead of $\omega_{TF}$, as we did in Fig.~\ref{spectroscopy}, since $\omega_{TF}$ depends on $a_f$. We vary the loss as a function of scattering length according to Eq.~\eqref{eq:loss}. The red dashed line reproduces the solid blue curve from Fig.~\ref{spectroscopy}. We additionally show the spectrum for a smaller final scattering length $a_f=3000a_0$ (blue dotted curve). Compared to the $a_f = 5000a_0$ spectrum, the minimum here shifts to larger $\omega_i$ as the final gas is narrower. The peak here is much sharper at small scattering length. Due to the lack of appreciable loss, the majority of dynamics are due solely to mode-mismatching. 

The yellow dashed-dot curve shows a much larger scattering length when $a_f = 10000a_0$. Here, the loss is so significant that the condensate is almost overdamped. Thus, there is near exponential decay regardless of the initial frequency. The spectroscopic feature becomes more shallow and broader, but is nevertheless still resolvable. At this large $a_f$, our LHY model is likely not fully describing the gas, as now $na^3 = 0.132$. It is interesting to consider exactly how the location of the minimum shifts as $a_f$ is varied. 

\begin{figure}[tb]
  \centering
    \includegraphics[width = 0.48\textwidth]{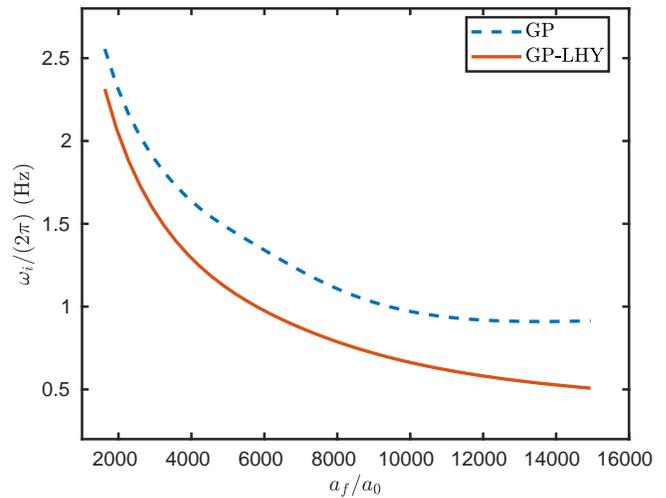}
    \caption{Optimum frequencies that minimize mode mismatch $\Delta$ for a range of final scattering lengths. Final trap is as in Table I, while the three-body loss is adjusted for scattering length according to Eq.~\eqref{eq:loss}. }
    \label{BeyondLHY}
\end{figure}

Figure~\ref{BeyondLHY} shows the initial frequencies $\omega_i = \omega_{opt}$ that give the minimum value of $\Delta$ as a function of final scattering length $a_f$, for two different models. The dashed blue curve gives the initial frequency predicted by a Gross-Pitaevskii (GP) model neglecting the LHY correction, which is contrasted with the red curve that shows the initial frequency predicted by the full model (GP + LHY). Deviation from the value predicted by the GP curve would constitute a measurement of the LHY correction. This difference is fairly subtle for small scattering length, but becomes more pronounced around $4000a_0$. Furthermore, we show these curves up to $14000a_0$ where the  beyond LHY correction is sizable, as $na^3 = 0.295$. Then, the difference between the optimal $\omega_i$ and the value predicted by the full (GP + LHY) model could give a measurement of the next, beyond LHY correction to the equation of motion.

In addition, based on the way in which it is selected, the initial density profile approximately measures the density profile of the strongly-interacting ground state. When the density profile is thought to be known exactly, shaping the trap, as discussed in Sec.~\ref{trapshaping}, could confirm if this is the actual ground state density.

One could also measure the ground state energy by a slightly modified quench experiment, especially in the intermediate regime where an adiabatic ramp is not possible but the gas is fully described by some (possibly modified) GP equation and loss is low. First, the density profile of the strongly-interacting Bose gas in some final trap needs to be established either theoretically or by finding the minimum of $\Delta$. Then, one could adiabatically prepare the weakly interacting gas in a state with density that matches the density of the final state, either in a harmonic or shaped initial trap. Measurement of the energy of this final state could be done by quenching from $a_i \rightarrow a_f$, while simultaneously turning off the trap. The final trap never actually needs to be produced. At small times, once the scattering length has been quenched, there is a strongly-interacting ground state for the final trap, even in the absence of that trap. The gas then expands, and as long as three-body loss is negligible during the expansion, the energy can be extracted from the final kinetic energy.

This energy measurement presumes, as we have presumed throughout, that the quench to the final state is adiabatic with respect to quasiparticles and molecular dynamics. That there is a quench rate that satisfies this criteria was argued in Sec. I. 
However, it is also interesting to contemplate quenches that are chosen to be diabatic with respect to two-body physics, to emphasize the creation and dynamics of quasiparticles. If this quench occurs within the mode-matching procedure, the resulting dynamics will be dominated by quasiparticles, disentangled from collective oscillations in the density profile of the gas.
\section{Outlook}

The mode-matching protocol represents a compromise between the need to get {\it quickly} to the final state, versus the need to place the gas {\it gently} into this state. As such, it represents a potential tool for probing BEC further beyond the mean-field regime than has previously been done. For modestly beyond mean-field gases that are still expected to be described by LHY theory, the results of a mode-matching experiment can be predicted accurately. The same experiment could be performed even well beyond this region, revealing something new. It is worth noting that, because of the ability to prepare initial BECs of arbitrary density profile as described in Sec.~\ref{trapshaping}, a similar protocol could be carried out in box-shaped traps \cite{gaunt2013bose}.

It would be tempting to apply the same type of measurement to the unitary gas whose scattering length is infinite. Thus far, however, the unitary Bose gases that have been produced decay on time scales short compared to trap periods. In such a case, the density variation versus time in Fig. \ref{densityDynamics} would be overdamped, and the figure of merit $\Delta$ poorly determined. This is not to say that the dynamics will not vary for quenches from various initial traps $\omega_i$; however, it would be difficult to predict or interpret the results of these experiments at present. To do so would require a hardy dynamical theory of the unitary Bose gas. Such models would, however, make concrete predictions for the density profile of the gas. This profile can be prepared in the initial state, whereby the results of projection would presumably reveal details of the gas beyond its shape. 

\section*{Acknowledgements}
This material is based upon work supported by the National Science Foundation under Grant Number PHY 1125844 and Grant Number PHY 1806971.

\bibliographystyle{apsrev4-1} 
\bibliography{UnitaryBIB} 
\end{document}